\documentstyle[11pt,newpasp,twoside,epsf,fleqn]{article}
\def\edcomment#1{\iffalse\marginpar{\raggedright\sl#1\/}\else\relax\fi}
\reversemarginpar

\newcommand{\gcs}{g\,cm$^{-2}$s$^{-1}$}
\newcommand{\gc}{g\,cm$^{-2}$}
\newcommand{\msun}{M$_{\sun}$}

\begin{document}

\title{Radiation-hydrodynamic Models of the Accretion Spots in Magnetic
Cataclysmic Variables} \author{K. Beuermann}
\affil{Universit\"ats-Sternwarte G\"ottingen, Geismarlandstr. 11,
D-37083 G\"ottingen, Germany}

\begin{abstract}
The structure of the near-polar accretion spots on accreting magnetic
white dwarfs has been studied theoretically and observationally in
numerous papers over the last decade. Detailed treatments are
available for the regime of low mass flux, usually termed the
bombardment case, and for higher mass fluxes which create a strong
shock standing above the photosphere of the white dwarf. No general
treatment is so far available for the case of shocks buried deep in
the photosphere. \mbox{I review} the theoretical foundations, present some
applications of theory, and discuss in short the open questions which
still need to be addressed. \end{abstract}

\section{Introduction}

Observational evidence indicates that the near-polar accretion regions
on the white dwarfs at least in some magnetic CVs (mCVs) are highly
structured, displaying rapid fluctuations of the mass flux both in the
spatial coordiates and in time. Since information of their structure
is lacking, a look at well-studied cases of magnetic accretion in the
solar system may be inspiring: (i) Jupiter's auroral oval displays
substantial structure in its UV emission as observed with the Hubble
Space Telescope, and (ii) the Earth's aurora shows narrow striations
which extend along field lines and indicate different densities in
adjacent flux tubes.  Although the physics is different, similar
structures {\it may} exist in mCVs.

Present models of mCV accretion spots are not sufficiently
sophisticated to account for such structure. They assume a spatially
constant mass flux across the column and are respresentative rather of
an individual subcolumn than of the accretion region as a whole. The
radiative interaction between individual subcolumns represents a
complication which is usually avoided. This neglect may be justified
in the optically thin case, but represents a severe restriction if the
column is optically thick.  In this review, I will summarize the
efforts made so far in understanding the properties of accretion
columns with special emphasis on the solution of the coupled
hydrodynamic and radiative transfer equations.

\begin{figure} 
\plottwo{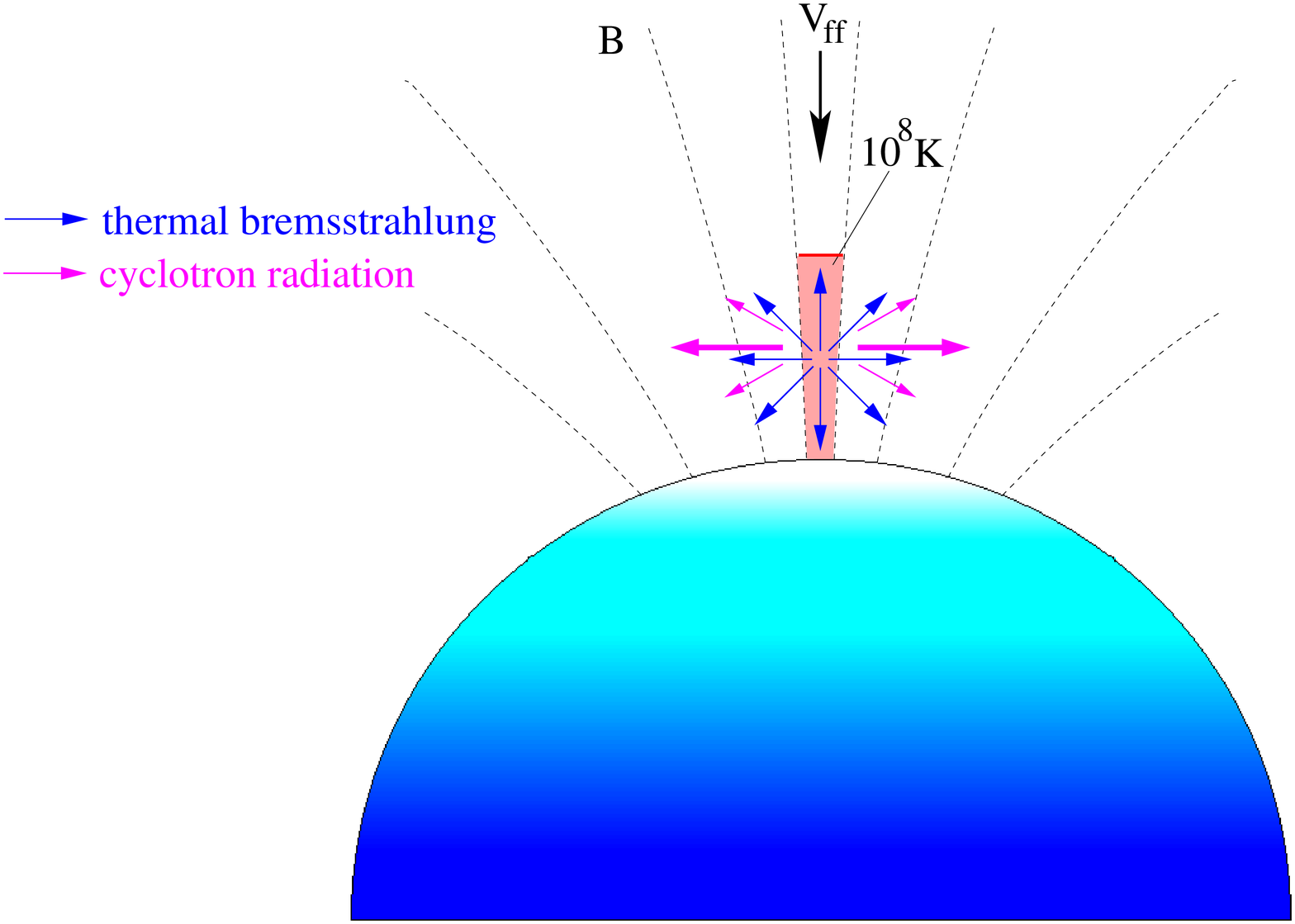}{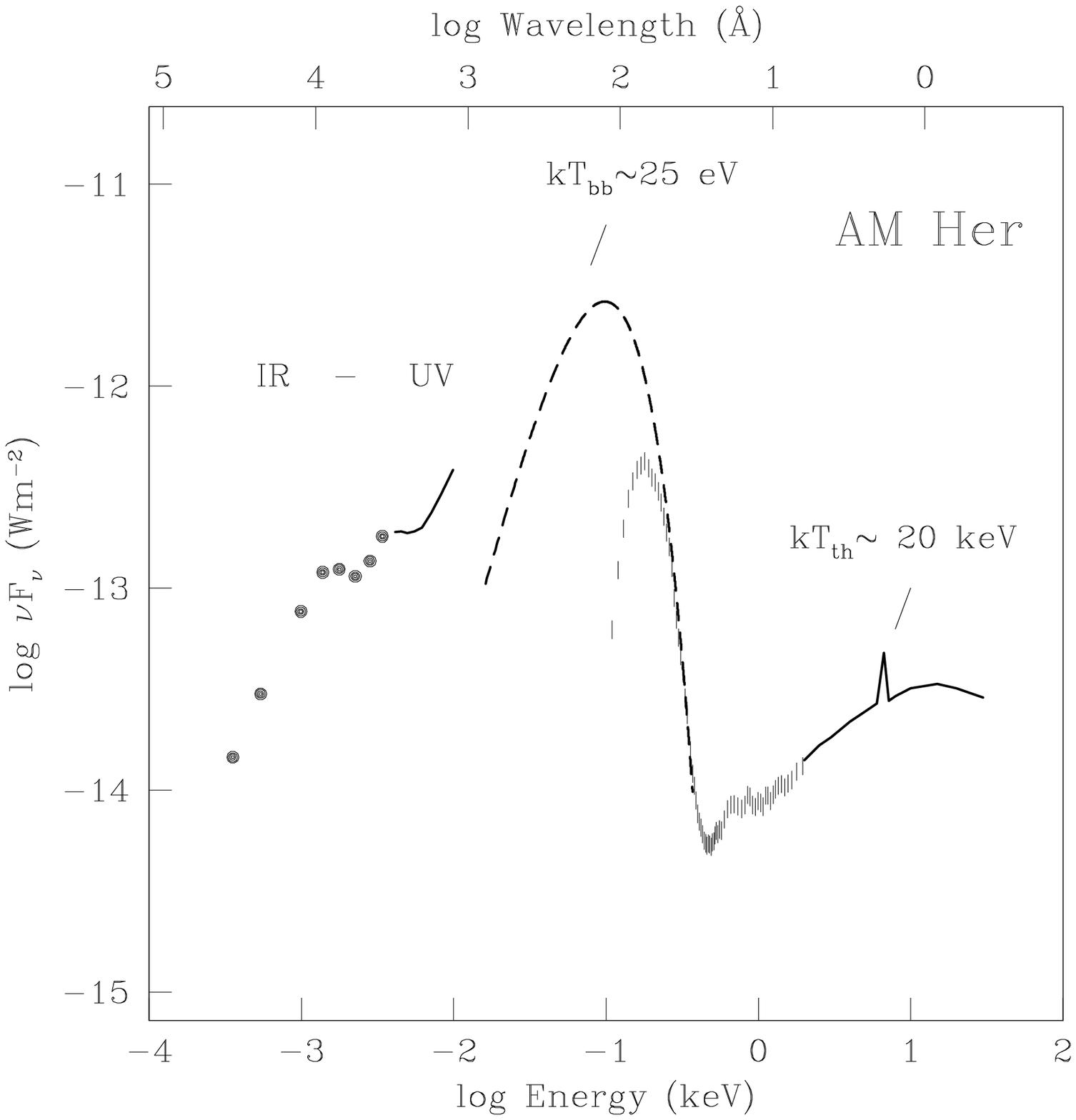}
\caption{{\it (a) Left: } Schematic diagram of a (tall) accretion
region on the white dwarf (from G\"ansicke, priv. comm.). {\it (b)
Right: } Overall spectral energy distribution of the prototype polar
AM Herculis at orbital phase $\phi \simeq 0.4$. The dots represent the
UBVRIJHK fluxes, the soft X-ray spectrum is from the April 1991 ROSAT
observation, and the IUE UV and the HEAO-1 hard X-ray spectrum are
shown by solid lines.  The dashed line represents the unabsorbed soft
X-ray \mbox{(quasi-)blackbody spectrum}.}
\end{figure}

\section{The Accretion Scenario}

Figure 1a schematically shows the pole cap of the magnetic accreting
white dwarf in a polar. The accretion column is fed by a mass flux
$\dot m = \rho_{\rm o}v_{\rm o}$ (cgs unit \gcs) with $\rho_{\rm o}$
the density and $v_{\rm o}$ the value of the frre-fall velocity
$v_{\rm ff}$ at the shock front.  The size of the accretion spot
depends on the coupling processes in the L$_1$-point and in the
magnetosphere of the white dwarf, while the stand-off distance of the
shock front, $h \simeq \frac{1}{4}\,v_{\rm o}t_{\rm cool}$ with $t_{\rm cool}$ the
cooling time scale, depends on the emission process, i.e. on
$\rho_{\rm o}$, the electron temperature $T_{\rm e}$, and the magnetic
field strength $B$. Depending on the parameters, the emission region
in mCVs is either tall (as shown in Fig. 1) or pill-box shaped
(i.e. smaller than wide) and the dominant emission process is either
hard X-ray bremsstrahlung or IR/optical cyclotron
emission. Reprocessing in the atmosphere of the white dwarf produces,
in addition, soft X-ray emission from the immediate surrounding of the
impact region and UV emission from a wider irradiated area.

The fact that some polars are dominated by soft X-ray emission has
been discussed in the literature for some 20 years and dubbed the
``soft X-ray problem''. In the original model by Lamb \& Masters
(1979) and King \& Lasota (1979), the white dwarf was assumed to
intercept $\la 50$\% of the emission, i.e. the non-reflected fraction
of the downward flux, and re-radiate it as soft X-rays. Kuipers \&
Pringle (1982) effectively solved the mystery of the intense soft
X-ray emission by noting that dense blobs of matter would penetrate
the photosphere and the subsequent optically thick radiative transfer
would reprocess all emissions into quasi-blackbody soft X-rays (see
alo Frank et al. 1988).

As an example, Figure 1b displays the overall spectral energy
distribution of the prototype polar AM Herculis at an orbital phase at
which near-maximum cyclotron and X-ray emission is seen. The spectral
energy distribution nicely displays the expected components: (i)
cyclotron emission at infrared and optical wavelengths; (ii) the
heated polar cap of the irradiated white dwarf in the UV longward of
the Lyman edge; (iii) an intense soft X-ray component; and (iv) a hard
X-ray bremsstrahlung (+ emission line) component. The companion star
and the unheated fraction of the white dwarf are comparatively faint
light sources. All accretion-induced emissions fluctuate in
time. E.g., the soft X-ray flux of AM Herculis in its high state
displays statistically relevant variations down to a time scale of
$\sim 0.2$\,s, as shown by ROSAT PSPC observations. This time scale
corresponds to density fluctuations in the free-falling pre-shock
accretion flow on length scales as short as $10^8$\,cm. An earlier
interpretation of the EXOSAT soft X-ray light curve of AM Her (Hameury
\& King 1988) suggested already that many such intependent subcolumns
are present at any time. For simplicity, most models disregard the
inhomogeneities in the pre-shock flow in space and time.

For sufficiently high mass flux, a standing shock separates the
infalling matter from the shocked plasma which slowly cools and
finally settles onto the white dwarf. An analytic solution for the
structure of an optically thin postshock flow cooling by
bremsstrahlung was given by Aizu (1973). If the flow is optically
thick, which is always the case for cyclotron emission in the
fundamental, the solution becomes much more complicated because the
radiative transfer has to be treated simultaneously with the
hydrodynamics (Woelk \& Beuermann 1996). 

\section{Bombardment solution}
 
In a hydrodynamic shock, the ions are stopped by Coulomb collisions
and the shock thickness equals about one ion mean free path. A special
situation arises if cooling is so strong that the randomized energy
can be radiated by the electrons within that scale length, i.e.,
effectively ``on the spot''. In this case, an {\it extended} post-shock
cooling flow does not exist and the concept of a discontinuity
separating two regimes of the flow is no longer appropriate. The whole
flow has now collapsed to a width which corresponds to the shock
thickness and effectively can be described as a hot corona forming
part of the white dwarf atmosphere. This mode is referred to as the
bombardment solution.

The required efficient cooling for this case can be provided by
cyclotron radiation in a strong magnetic field which causes the
electron temperature to stay far below the Rankine-Hugoniot 
temperature $T_{\rm sh}$ of the shock solution. Originally suggested
by Kuijpers \& Pringle (1982), this case was rigorously solved by
Woelk \& Beuermann (1992). They treated the ion collisional energy
loss by means of statistical plasma physics and simultaneously
accounted for the losses of the electrons by a fully
frequency-dependent and angle-dependent radiative transfer in a
plane-parallel heated atmosphere. The solution predicts a peak
electron temperature which increases as k$T_{\rm e} \propto \dot
m^{0.42}$ (Woelk \& Beuermann 1993, see also Fischer \& Beuermann
2001) and equals 1\,keV for $\dot m \simeq 0.01$\,\gcs, $B \simeq
30$\,MG, and a white dwarf mass of 0.6\,\msun. Such low temperatures
are observed in AM Her stars with very low accretion rates (Schwope
this conference) and suggest that the predicted temperatures are
roughly correct. The solution becomes invalid when $T_{\rm e}$
approaches the Rankine-Hugoniot  temperature with increasing $\dot m$.

\section{Radiation-hydrodynamics}

For $\dot m \ga 0.1$\,\gcs, the post-shock plasma flow and the
radiative transfer have to be treated simultaneously. The only
feasible approach is a fluid description which does not allow to
resolve the shock front, however. There is no simple connection
between such an approach and the kinetic description of the
bombardment solution. Nevertheless, we find that the fluid
approach for low $\dot m$ recovers some important apects of the
bombardment case.

\subsection{The Equations}

The general case assumes a two-fluid plasma and allows for different
temperatures $T_{\rm i,e}$ and bulk velocities $v_{\rm i,e}$ of ions
and electrons. At the high densities in mCV shocks, Coulomb collisions
will assure $\vec{v}_{\rm i} = \vec{v}_{\rm e} \equiv \vec{v}$, except
perhaps within the (unresolved) shock front.  A two-fluid treatment
may then proceed with a single equation of motion which contains the
total pressure, i.e. the sum of ion pressure $P_{\rm i}$ and electron
pressure $P_{\rm e}$. The skalar pressure and the temperature are
related by $P_{\rm i,e} = n_{\rm i,e}kT_{\rm i,e}$ (with $n_{\rm i}$
and $n_{\rm e}$ the particle densities of ions and electrons), which
assumes that the particle distributions are isotropic and
Maxwellian. Cyclotron losses could, in fact, produce an anisotropic
electron distribution function, but the optical depth in the
fundamental harmonic is so large that radiative transfer causes the
temperatures perpendicular and parallel to the magnetic field to agree
closely, except very near the surface of the emission volume
(Zhelezniakov 1983).  For simplicity, we assume a hydrogen plasma in
which $n_{\rm i} = n_{\rm e}$ and the mean molecular weight is $\mu =
1/2$.

The hydrodynamic equations include the continuity equation, the
equation of motion, and the energy equations for ions and electrons,
which describe the conservation of mass, momentum, and energy,
respectively:
\begin{equation}
\frac{d\rho}{dt} + \rho(\nabla\cdot\vec{v}) = 0 
\end{equation}
\begin{equation}
\frac{d\vec{v}}{dt} + \frac{1}{\rho}\nabla(P_{\rm i}+P_{\rm e}) 
= \vec{g} + \vec{f}_{\rm rad}
\end{equation}
\begin{equation}
\frac{dP_{\rm i}}{dt} + \gamma\frac{P_{\rm i}}{\rho}
\frac{d\rho}{dt} = -(\gamma-1)\left(\Lambda_{\rm ei}+\Lambda_{\rm h}\right)
\end{equation}
\begin{equation}
\frac{dP_{\rm e}}{dt} + \gamma\frac{P_{\rm e}}{\rho}
\frac{d\rho}{dt}=(\gamma-1)\left[\Lambda_{\rm ei}+\Lambda_{\rm h}- 
\nabla\cdot(\vec{q}+\vec{F}_{\rm rad})\right].
\end{equation}
Here, $\frac{d}{dt} = \frac{\partial}{\partial t}+\vec{v}\cdot\nabla$,
$\rho = m_{\rm u}n_{\rm i}$ is the mass density with $m_{\rm u}$ the
unit mass ($\simeq$ proton mass), and $\gamma = 5/3$ is the adiabatic
index. On the right hand sides, the equation of motion contains the
gravity $\vec{g}$ and the volume force exerted by absorbed and
scattered radiation, $\vec{f}_{\rm rad}$. The latter becomes important
when the accretion rate approaches the Eddington limit. In the energy
equations, $\Lambda_{\rm ei}$ describes the energy exchange between
ions and electrons by Coulomb collisons and $\Lambda_{\rm h}$ refers
to electron heating by any other process, with the energy provided
again by the ions. Additional energy gains of the electrons are described
by the divergence of the heat conduction flow $\vec{q}$ and of the
radiative flux $\vec{F}_{\rm rad}$. Calculation of the radiative flux
(Eq.\,5) requires solution of the radiative transfer equation (6) for
the intensity $I_{\nu}(\vec{n},\nu)$ as a function of the direction
$\vec{n}$ and the frequency $\nu$, where $\sigma$ and $\chi$ are the
scattering and absorption coefficients, $J_{\nu}$ is the angle-averaged
intensity and the use of $B_{\nu}$ for the source function indicates
the assumption of LTE:
\begin{equation}
\vec{F}_{\rm rad} = \int\limits_{0}^{\infty}\oint\limits_{4\pi} 
I_{\nu}\,\vec{n}\,d\omega \,d\nu,
\end{equation}
\begin{equation}
\vec{n}\cdot\nabla I_{\nu} = - (\sigma + \chi)I_{\nu} 
+ \chi B_{\nu} + \sigma J_{\nu}.
\end{equation}
Unsurprisingly, Eqs.\,(1)--(6) have not been solved in all their
beauty, but only after the introduction of some rather severe
simplifications.

\subsection{Simplified Geometry}

The correct funnel geometry (see Canalle, this conference) is most
complicated so solve, the spherical or the linear geometry represent
the simpler cases of which the latter is an acceptable approximation
only for shock heights $h_{\rm sh} \ll R_{\rm wd}$.  Realistic columns
are always limited in the direction perpendicular to the flow. This
limitation is of no concern when the radiative losses can be treated
as optically thin emission, but in the optically thick case it
requires a multi-dimensional radiative transfer (i.e., with the {\it
direction} of $\nabla T_{\rm e}$ position-dependent).  Comparatively
simple cases are (i) a shallow infinite layer with the radiative flux
vector constantly pointing in the radial direction (Woelk \& Beuermann
1996, Fischer \& Beuermann 2001) or (ii) a tall column with the
radiative flux vector pointing only sideways (Cropper et
al. 1999). Naturally, gravity effects increase with column height.

\subsection{Electron Heating by Non-Coulomb Processes}

In Eqs.\,(3) and (4), the quantity $\Lambda_{\rm h}$ represents any
heating process other than Coulomb collisions. Non-Coulomb heating of
electrons is known to occur in interplanetary shock fronts, in
particular, the bow shock of the Earth (e.g. Stone \& Tsurutani
1985). In these collisionless shocks (with a typical particle density
of $n\sim 10$\,cm$^{-3}$) characteristic processes are the reflection
of ions, charge separation between electrons and ions, and the
incidence of various types of plasma instabilities which lead to rapid
electron heating. Plasma instabilities may also play a role in heating
the electrons in supernova shocks (Lesch 1990). Whether such processes
occur in mCV shocks with $\vec{v} \parallel \vec{B}$, high Mach
numbers, and high particle densities (exceeding those in
interplanetary space by factors of $10^{11}$ to $10^{16}$) needs to be
studied. Some authors (e.g., Saxton \& Wu 2001) assume the
presence of plasma heating in the shock front and consider $T_{\rm e}$
immediately behind the shock a free parameter, possibly ranging up to
$T_{\rm i}$. Others, like Woelk \& Beuermann 1992, 1996) and Fischer
\& Beuermann (2001) have disregarded plasma heating and considered
Coulomb collisions only ($\Lambda_{\rm h} = 0$). In their solutions,
the ion temperature displays a (quasi-)discontinuity at the shock,
while the electron temperature increases only on a length scale which
equals the Coulomb time scale times the flow speed. Clearly, the
question of plasma heating in the shock front needs further
consideration.

\subsection{One-fluid Plasma}

Electron and ion temperatures will be similar if the time scale for
energy exchange between ions and electrons by Coulomb collisions is
short compared with the cooling time scale. A one-fluid plasma is a
reasonable approximation for high $\dot m$ and low $B$, while a
two-fluid approach is required for low $\dot m$ and high $B$. The
dividing line between the two regimes is approximately located at
$\dot m B_7^{-2.6} \simeq 0.1$, where $\dot m$ is in \gcs\ and $B_7$
is in units of $10^7$\,G (see Fig.\,4 below). Hence, $T_{\rm e} \simeq
T_{\rm i}$ for $\dot m B_7^{-2.6} \gg 0.1$ and $T_{\rm e} < T_{\rm i}$
for $\dot m B_7^{-2.6} \ll 0.1$.

\subsection{Cooling functions}

The free-free optical depth of the post-shock flow parallel to
$\vec{v}$ is $\tau_{\rm ff} \simeq 0.1\ldots1.0$. Hence, a flow
cooling by bremsstrahlung only is optically thin to a first
approximation with $\nabla\cdot\vec{F}_{\rm rad} = 1.4\,10^{-27}n_{\rm
e}^2T_{\rm e}^{1/2}$\,erg\,cm$^{-3}$s$^{-1}$. In this case, the
divergence term is a locally defined quantity which obviates the need
for solving the radiative transfer equation. Most authors have
simplified the radiative transfer by replacing
$\nabla\cdot\vec{F}_{\rm rad}$ by a generalized cooling function
expressed as the sum of terms of the form $\rho^aT_{\rm e}^b \propto
\rho^{a-b}P_{\rm e}^b$ with $a, b$ fixed for each term. Bremsstrahlung
is represented by $a = 2, b = 0.5$, cyclotron radiation by $a = 0.15,
b = 2.5$ (Saxton \& Wu 2001, and references therein). The concept of a
cooling function is applicable in approximately the same limit as the
one-fluid plasma, $\dot m B_7^{-2.6} \gg 0.1$ (see Fig.\,4 below).
This limitation applies to radiative transfer through the shock
front. For tall columns which lose energy preferentially through their
sides, cooling functions are applicable also for 
lower values of $\dot m B_7^{-2.6} $ 

\subsection{Energy Transport by Conduction}

The heat conduction flow, $\vec{q} \propto n_{\rm e}T_{\rm
e}^{5/2}\nabla{T_{\rm e}}$, is of importance only in regions of high
particle density, high electron temperature, and/or a large gradient
of the latter, i.e. (i) in the transition between the cooling flow and
the stellar atmosphere and (ii) in the shock front with the adjoining
pre-shock flow (Imamura et al. 1987). Most attempts to solve
Eqs.\,(1)--(6) proceed with $\vec{q} = 0$. Wu (2000) has summarized
the effects of heat conduction in mCVs.

\subsection{Atomic Line Emission}

The post-shock region is optically thick for the stronger atomic
emission lines, complicating their calculation. One may adapt a
procedure known from stellar atmosphere research which involves two
steps: (i) the temperature structure is calculated assuming continuum
absorption and emission only; (ii) the emerging emission line fluxes
are calculated using this temperature structure. Tennant et al. (1998)
and Wu et al. (2001) have used a cooling function for step (i)
and a simplified line transfer for step (ii). An accurate calculation
of the line profiles and line intensities requires substantial effort.

\section{Oscillatory Solutions and Linear Stability Analysis}

Time-dependent solutions of the hydrodynamic equations can be used to
study the stability of stationary solutions. The added complication of
the time dependence is compensated for by the simplicity of the
coolong function. The stability analysis proceeds by, e.g., (i)
assuming small perturbations of the position of the shock front and of
the plasma parameters, (ii) linearization, and (iii) analysis of the
growth rates of the harmonics of the oscillations (Langer et al. 1982,
Saxton \& Wu 2001, Wu 2000 and references therein). The possible
frequencies are the equivalent to the harmonic oscillations of a pipe
which is open at one end.

Saxton \& Wu (2001) confirm earlier findings that shocks cooling by
brems\-strahlung only are unstable against oscillations, except
possibly in the fundamental mode.  Shocks dominated by cyclotron
cooling, on the other hand, are stable in the fundamental and all
reasonably occurring higher oscillation modes.  This suggests that
stationarity is a reasonable model assumption if cyclotron emission is
the dominant cooling agent. The low-amplitude oscillations with
periods of $\sim 1-3$\,s seen in several mCVs (e.g. Larsson 1982)
probably originate from regions within the accretion spot which are
dominated by bremsstrahlung.

Solutions of Eqs.\,(1)--(6) with a time-dependent mass flux $\dot m$
incident on the white dwarf do not exist to my knowledge.

\section{Stationary Solutions with Cooling Function}

With $\nabla\cdot F_{\rm rad}=\Lambda_{\rm c} \propto \rho^{\rm
a-b}P^{\rm b}$ (or a sum of such terms), only the hydrodynamic
equations have to be solved.  In the stationary case for linear or
spherical geometry, Eqs. (1)--(4) reduce to a system of coupled
ordinary (though non-linear) equations, in which all plasma parameters
depend only on the radial coordiate.

\subsection{One-fluid Plasma}

For a one-fluid plasma, $P=P_{\rm e}+P_{\rm i}$ and Eqs.\,(3) and (4)
may be combined to a single energy equation. The linear case is
simplest and applies if the shock height is $h_{\rm sh} \ll R_{\rm
wd}$.  With $f_{\rm rad}=0$, $\Lambda_{\rm h}=0$, $\vec{q}=0$,
$\gamma=5/3$, $g \simeq $\,const, $z=r-R_{\rm wd}$, and $v < 0$, the
hydrodynamic equations reduce to
 \begin{equation}
\frac{d}{dz}(\rho v) = 0
\end{equation}
\begin{equation}
v\frac{dv}{dz} + \frac{1}{\rho}\frac{dP}{dz} = -g
\end{equation}
\begin{equation}
v\frac{dP}{dz} - \frac{5}{3}\,v\,\frac{P}{\rho} \frac{d\rho}{dz} = 
-\frac{2}{3}\Lambda_{\rm c},
\end{equation} 
\noindent For the above cooling function $\Lambda_{\rm c}$, a
closed-integral solution has been given Wu et al. (1994). Because of
the one-fluid assumption, the solution is limited to the high $\dot
m$, low $B$ part of the $\dot m$--$B$ plane.
In the spherical case, the mass flux $\rho v$ in Eq.\,(7) and $g$ in
Eq.\,(8) are $r$-dependent.  The linear and spherical cases can be
solved analytically if cooling is by bremsstrahlung only (Aizu 1973).

A potentially important case is that of a narrow, tall column (width
$\ll$ height $\ll R_{\rm wd}$) which emits cyclotron radiation
primarily sideways ($\perp B$). The radiative flux vector is then
approximately in the $x,y$-plane which effectively decouples
hydrodynamics and radiative transfer. With certain simplifying
assumptions on the transverse radiative transfer, the radiative losses
can be approximated by a cooling function (Wu et al. 1994, Cropper et
al. 1999).

\subsection{Two-fluid Plasma}

For strong cyclotron cooling ($\dot m B_7^{-2.6} \ll 0.1$), the
one-fluid approximation becomes invalid.  Electron and ion
temperatures differ, and Eq.\,(9) has to be replaced by
\begin{equation}
v\frac{dP_{\rm i}}{dz} - \frac{5}{3}\,v\,\frac{P_{\rm i}}{\rho}
\frac{d\rho}{dz} = -\frac{2}{3}\,(\Lambda_{\rm ei}+\Lambda_{\rm h}) 
\end{equation}
\begin{equation}
v\frac{dP_{\rm e}}{dz} - \frac{5}{3}\,v\,\frac{P_{\rm e}}{\rho}
\frac{d\rho}{dt} = \frac{2}{3}\left(\Lambda_{\rm ei} + 
\Lambda_{\rm h} - \Lambda_{\rm c}\right).
\end{equation}
Although the temperatures deduced this way will be more reliable, the
question of electron heating in the shock ($\Lambda_{\rm h}$) implies
a remaining uncertainty. Temperature profiles of ions and electrons
with different assumptions on $\Lambda_{\rm h}$ (and $\rho_{\rm o}$)
are depicted, e.g., in Saxton \& Wu (2001, their Fig. 3). For strong
heating, the profiles are characterized by a maximum at the shock and
a rapid decrease thereafter, while weak shock heating causes the
electron temperature to assume a maximum further downstream,
approaching the case of pure Coulomb heating.

\section{Stationary Plane-parallel Two-fluid Radiation-hydrodynamics}

The only truely radiation-hydrodynamic solution is that of Woelk \&
Beuermann (1996). Fischer \& Beuermann (2001) present further results
including a parametrization of the temperature profiles.

Their approach involves the solution of the four hydrodynamic
equations, Eqs. (7), (8), (10), and (4) with $\Lambda_{\rm h}=0$ and
$q = 0$
\begin{equation}
v\frac{dP_{\rm e}}{dz} - \frac{5}{3}\,v\,\frac{P_{\rm e}}{\rho}
\frac{d\rho}{dz} = \frac{2}{3}\left(\Lambda_{\rm ei} 
 - \frac{dF_{\rm rad}}{dz}\right),
\end{equation}
simultaneous with the radiative transfer equations (5) and (6). The
solution of the latter was developed from a Feautrier code for
plane-parallel stellar atmospheres, modified to include the anisotropy
of cyclotron absorption and emission. The radiative transfer is fully
frequency and angle dependent. The solution is energy conserving and
ensures that the radiative flux (Eq.\,6) exactly equals the accretion
energy $\frac{1}{2}\dot m v_{\rm o}^2$. However, because of the
plane-parallel ansatz the radiative flux is directed radially. In his
review, Wu (2000) incorrectly states that the diffusion approximation
was used and that energy is not conserved.

The solution relates $T_{\rm e}$ in a unique way to $\dot m$, $B$, and
$M_{\rm wd}$ as the physical input parameters. Fig. 2 shows examples
of the profiles of $T_{\rm e}(z)$, $T_{\rm i}(z)$, and $v(z)$ for
$B=30$\,MG and $\dot m=0.01, 0.1$ and 1.0\,\gcs. Electron and ion
temperatures are found to differ substantially near the shock front
and equilibrate only further downstream. All flows are optically thick
in the cyclotron fundamental which requires a negative electron
temperature gradient to allow the energy to escape through the shock
front. This is a consequence of the assumption of an {\it infinite}
layer and may no longer hold true for columns of finite width. The
effect of plama instabilities heating the electrons in the shock has
not so far been considered by us and one can only speculate as to the
effect they may have in the presence of optically thick radiative
transfer. If the radiation escapes through the shock front, radiative
transfer is likely to counteract plasma heating and will tend to
depress $T_{\rm e}$. I consider it unlikely that $T_{\rm e}$ at the
shock can be anything near the Rankine-Hugoniot temperature for high
$B$ and low $\dot m$.  The low temperatures of $\sim 1$\,keV observed
from some AM Her stars (e.g. Schwope, this conference) seem to support
this conclusion, but it is fair to state that the importance of plasma
instabilities in mCV shocks is far from understood and further study
is needed.

\begin{figure}
\plotfiddle{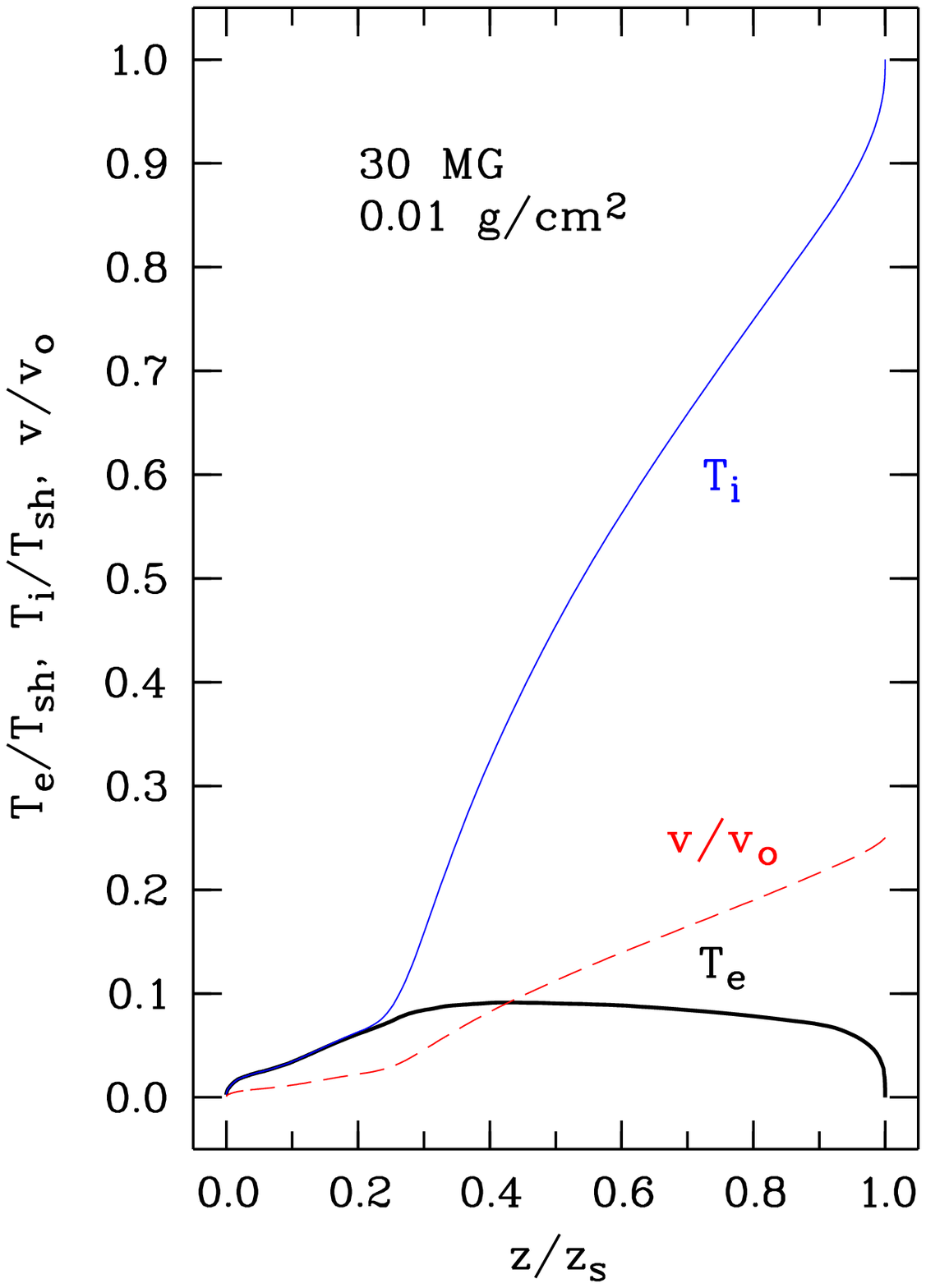}{13mm}{0}{33}{30}{-210pt}{-109pt}
\plotfiddle{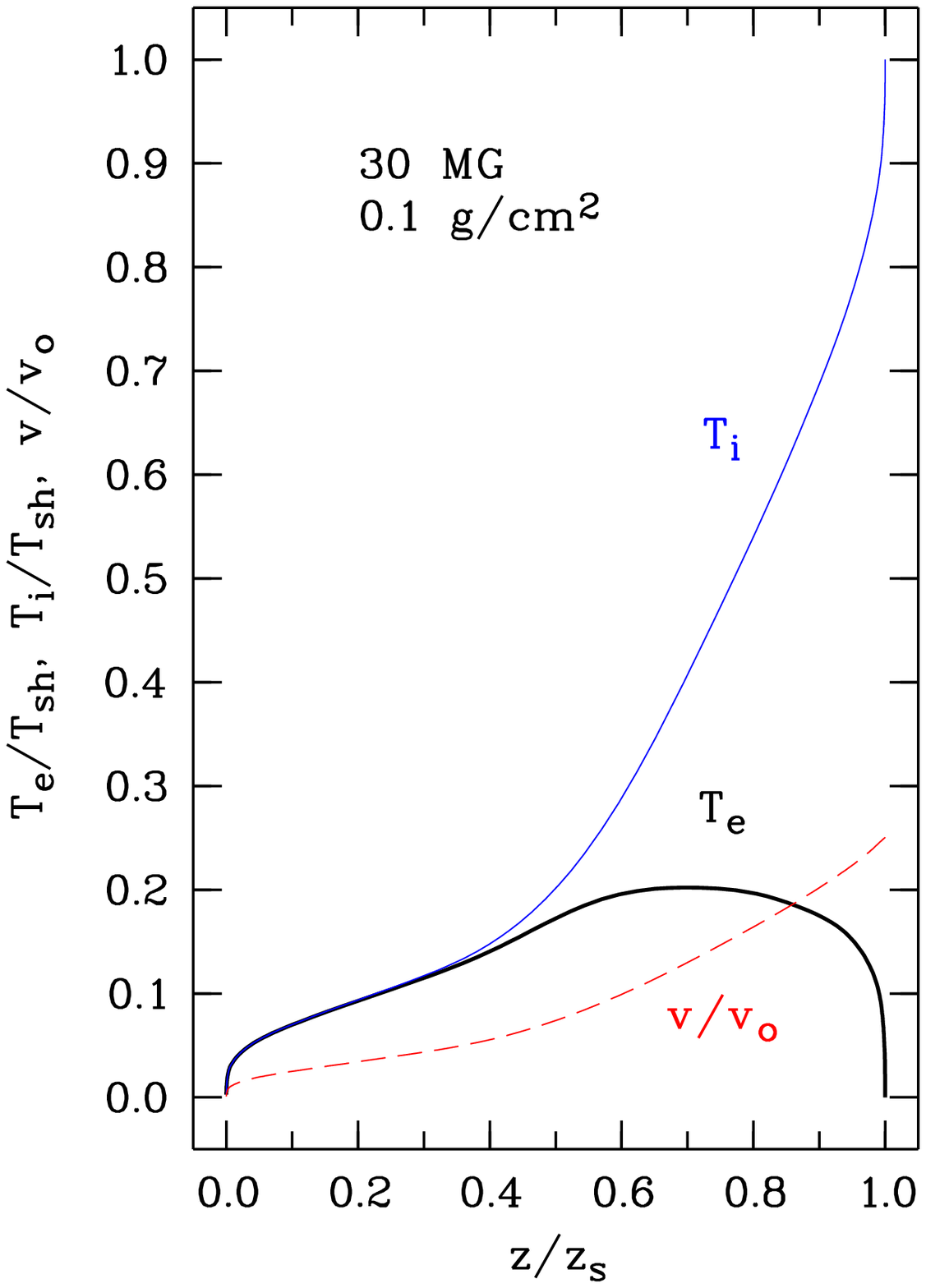}{13mm}{0}{33}{30}{-85pt}{-59pt}
\plotfiddle{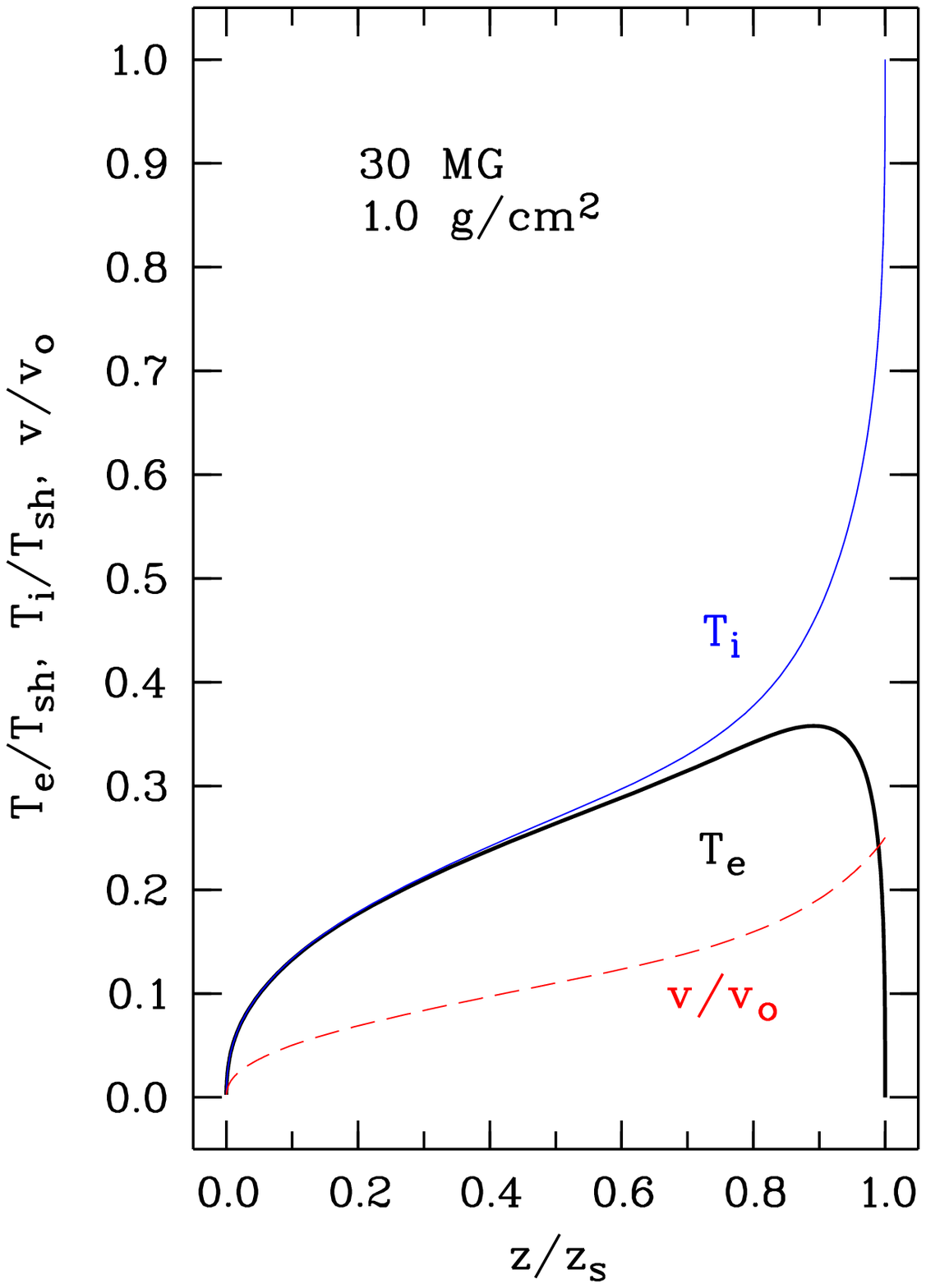}{13mm}{0}{33}{30}{40pt}{-9pt}
\caption{Profiles of $T_{\rm i},~T_{\rm e}$, and $v$ 
for $B=30$\,MG, and $\dot m = 0.01,~0.1,~{\rm and}~1.0$\,\gcs\ (from
Fischer \& Beuermann 2001).}
\vspace{5mm}
\plotfiddle{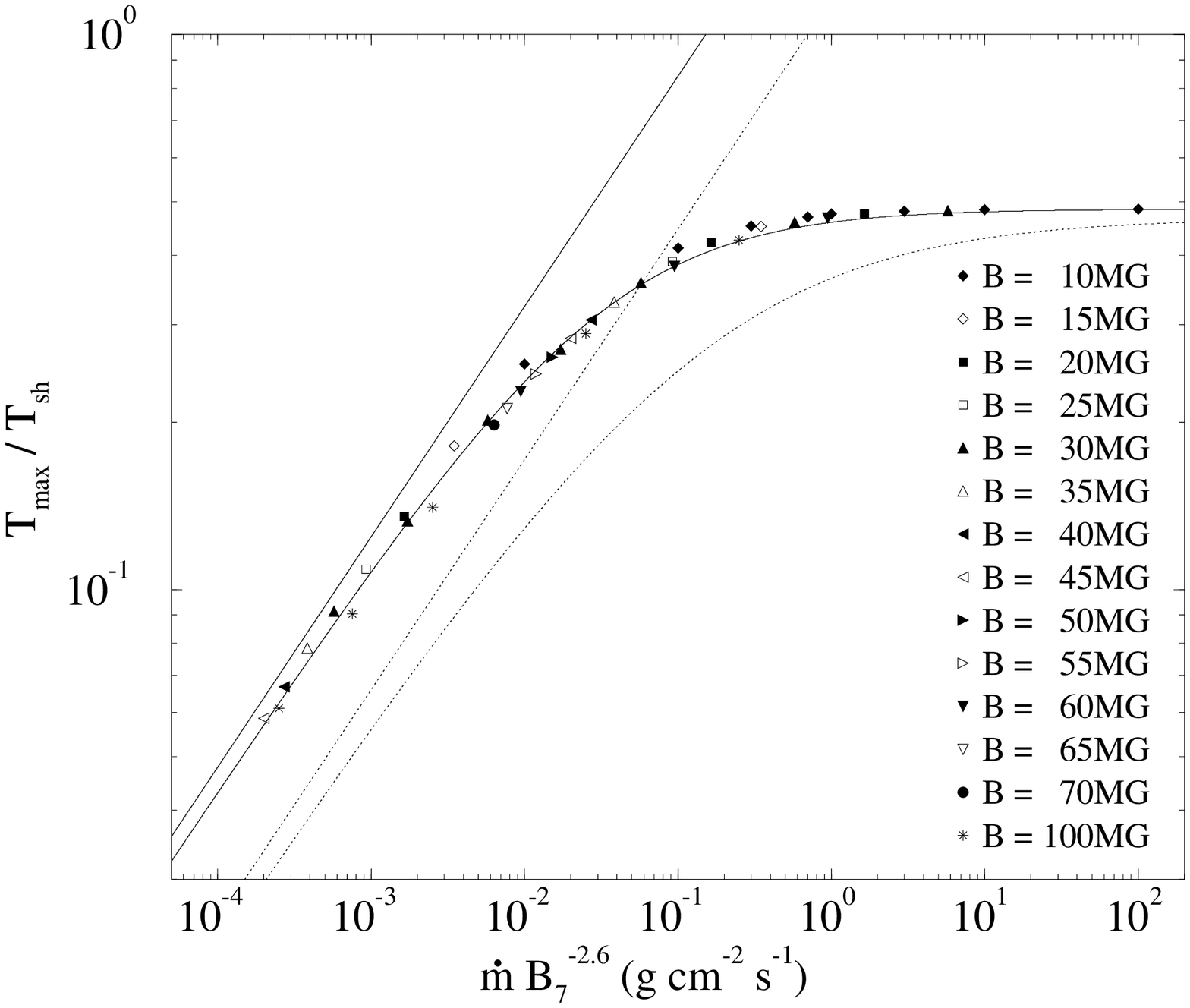}{22mm}{0}{33}{31}{-202pt}{-90pt}
\plotfiddle{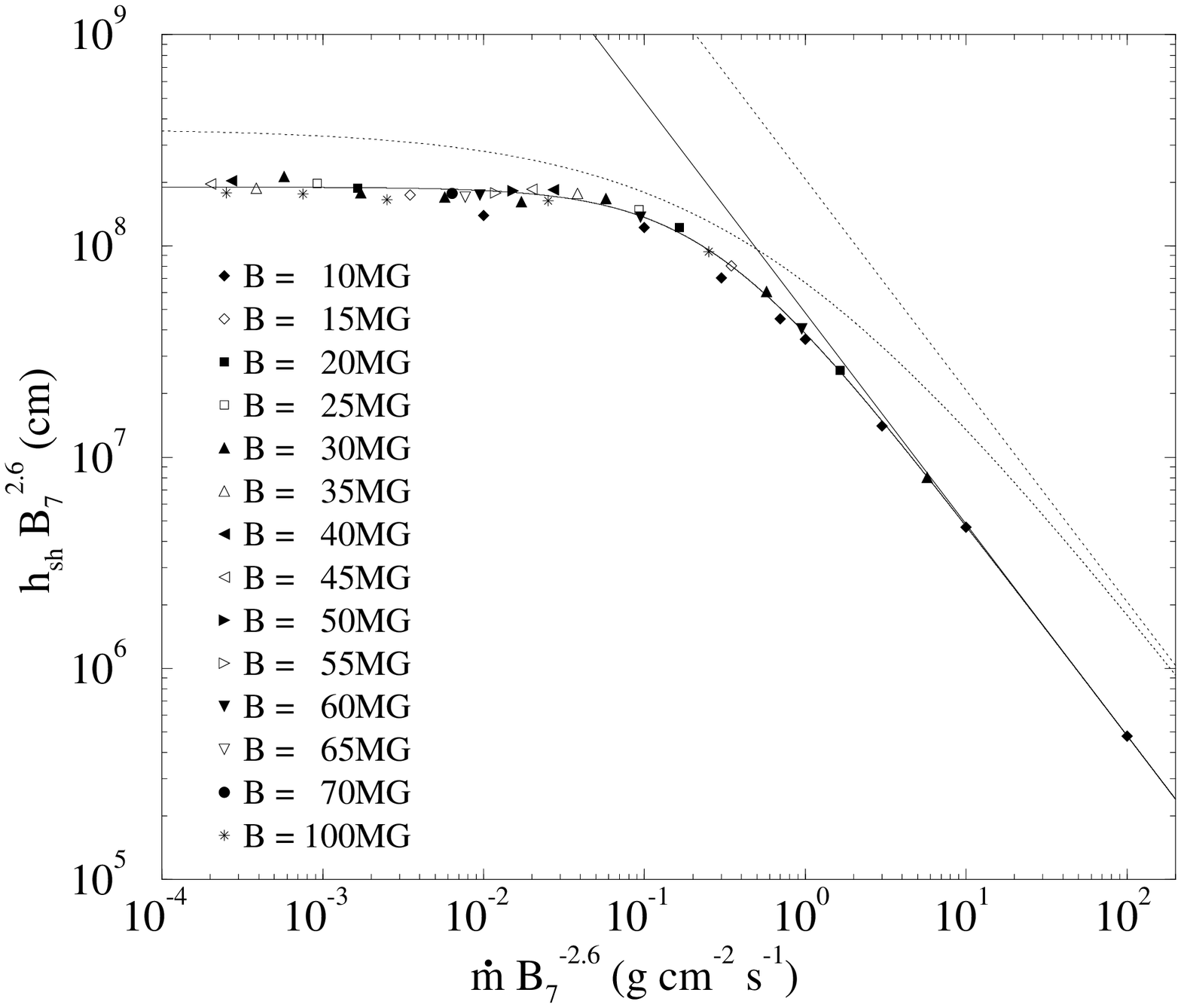}{22mm}{0}{33}{31}{-13pt}{-15xpt}
\caption{{\it (a) Left: } Peak electron temperature in units of the
ion temperature at the shock for various values of the field strength
vs. $\dot m B_7^{-2.6}$.  {\it (b) Right: } Normalized shock
height $h_{\rm sh} B_7^{2.6}$ vs. $\dot m B_7^{-2.6}$.}
\end{figure}

In Fischer \& Beuermann (2001), we have parameterized the profiles
$T_{\rm e}(z)$ by their peak value $T_{\rm max}$ and the shock height
$h_{\rm sh}$. Fig.\,3 shows $T_{\rm max}$ normalized to the ion
temperature at the shock which initially contains all the energy (left
panel) and a normalized version of the shock height (right
panel). Both representations unify the results for a large range of
model parameters (``data'' points) into a unique dependence on the
quantity $\dot m B_7^{-2.6}$ (``data'' points and solid curves for
0.6\,\msun, dotted curves for 1\,\msun). It is noteworthy that the
decreasing peak electron temperature for low $\dot m B^{-2.6}$ merges
into the results for the bombardment solution (straight lines). In the
same limit, the normalized shock height $h_{\rm sh}B_7^{2.6}$ (in cm)
becomes constant. From Fig.\,3 (right panel) one finds a limiting
shock height for the 0.6\,\sun\ white dwarf and for low $\dot m
B^{-2.6}$ of $h_{\rm sh} \simeq 1.9\,10^8\,B_7^{-2.6}$,
i.e. $10^7$\,cm for 30\,MG and as low as $5\,10^5$\,cm for
100\,MG. Since bremsstrahlung-dominated shocks at the same $B$ but
higher $\dot m$ will be still lower, the assumption of plane-parallel
or pillbox-shaped shocks is likely to be good for the higer field
strengths.

\begin{figure} 
\plotfiddle{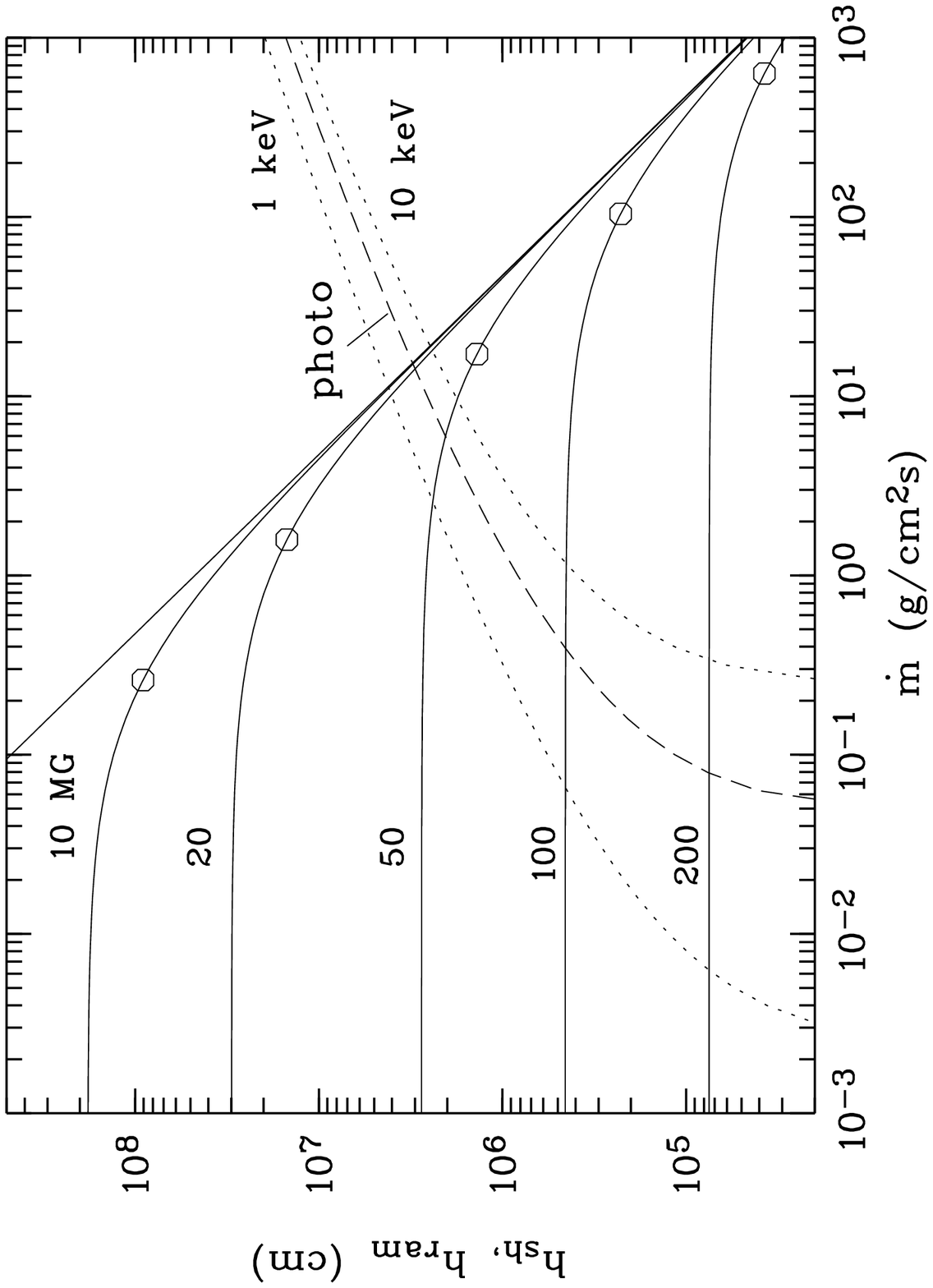}{60mm}{270}{45}{45}{-180pt}{250pt}
\caption{Shock height vs. mass flux $\dot m$ for a white dwarf mass of
0.6\,M$_{\sun}$ and five values of the polar field strength. Also
shown is the depression of the bottom of the post-shock flow below the
photosphere (photo) and the level in the undisturbed atmosphere
which corresponds to optical depths $\tau = 1$ for 1\,keV and
10\,keV photons (Beuermann \& Fischer in preparation).}
\plotfiddle{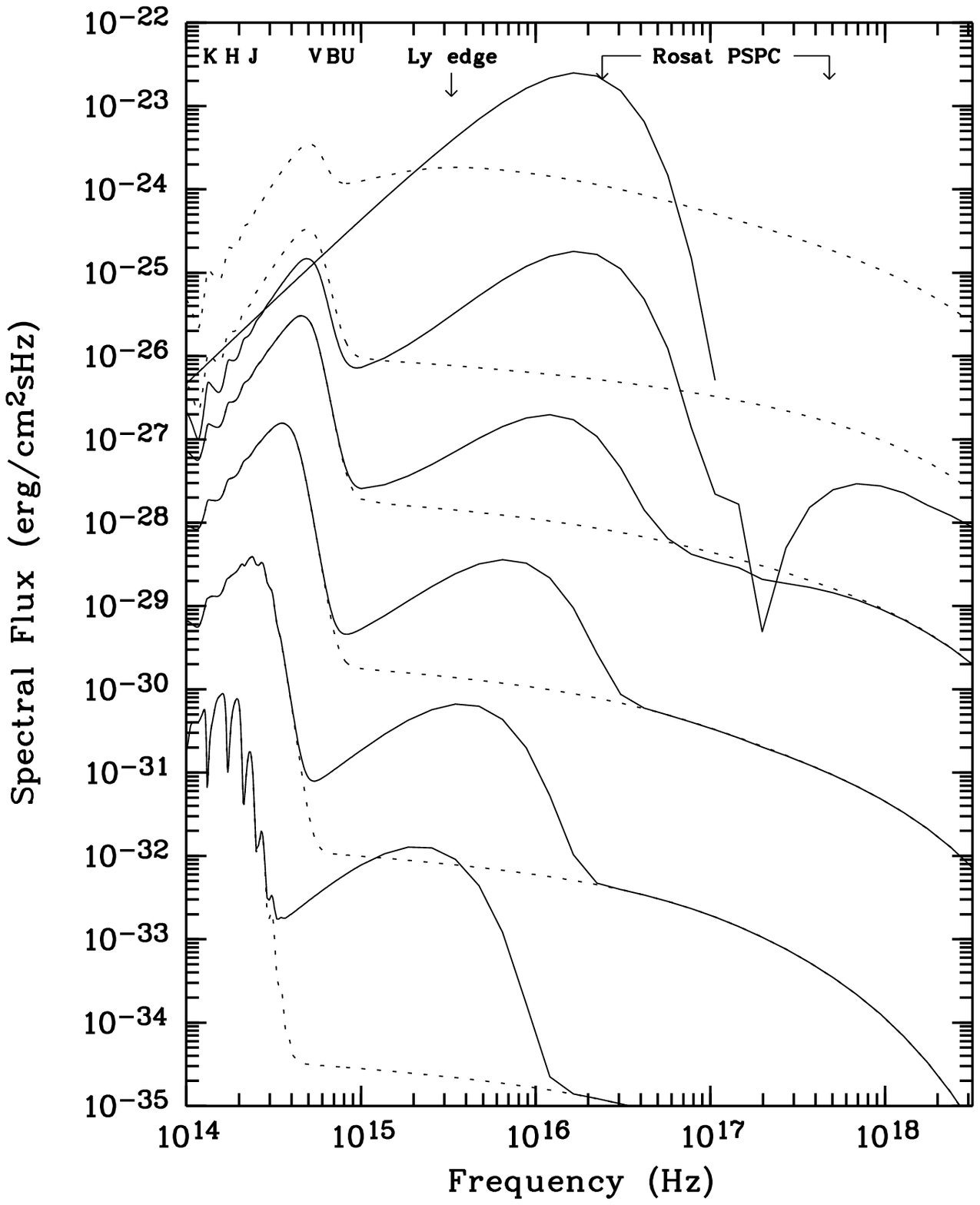}{87mm}{0}{52}{52}{-150pt}{-15pt}
\caption{Representative emerging spectra for $B=14$\,MG,
$\Theta=60^\circ$ against the field direction, and mass fluxes $\dot m
= 10^{-3},10^{-2},10^{-1},1, 10$, and 100\,\gcs, each for an emitting
area of $10^{15}$\,cm$^2$ at 100\,pc distance. Dashed curves are the
emitted spectra, solid curves the emerging spectra.  The blackbody
assumption is a crude approximation, in particular for the heated
white dwarf at low $\dot m$ (Beuermann \& Fischer in preparation).}
\end{figure}

In an attempt to generalize the results for a finite column width $D$,
we have calculated the additional energy losses by optically thick
radiation leaving their sides using a ray tracing technique (Fischer
\& Beuermann 2001). For a given $\dot m$, these added losses cause the
equilibrium temperature to be {\it lower} than in an {\it infinite}
layer. A correct treatment of the problem would require a truly 2-D
radiative transfer which presents a substantial complication and
introduces the with $D$ as an additional model parameter besides $\dot
m$. No such calculations are available in the literature. Fischer \&
Beuermann suggest a simple correction which ensures energy
conservation: the electron temperature derived for $\dot m_{\infty}$
for an infinite layer approximately applies to the higher value $\dot
m_{\rm D}=r \dot m_{\infty}$ for finite $D$, where $r={\cal L}/\dot m
v_{\rm o}$ is the ratio of the total radiative loss from the finite
region over the accretion energy (both per cm$^2$). They provide a
calibration of $r$ as a function of $h_{\rm sh/D}$ and $\dot m
B^{-2.6}$. Finally, note that in {\it all} models which approximate
the optically thick radiative transfer by an emergent Planck spectrum
cut off at some critical frequency $\nu^*$ energy is, by definition,
only approximately preserved.

\section{Applications}

\subsection{Free-standing vs. Buried Shocks}

The ram pressure $P_{\rm ram}=\rho_{\rm o}v_{\rm o}^2$ of the
infalling matter suppresses the bottom of the cooling flow 
into the atmosphere of the white dwarf. If also the shock front is
located within the atmosphere, the shock is considered submerged or
buried (Frank et al. 1988). Fig.\,4 compares $h_{\rm sh}$
vs. $\dot m$ (from Fig.\,3, right panel) to the depression produced by
$P_{\rm ram}$.  one finds that bremsstrahlung-dominated shocks are
buried for $\dot m \ga 10$\,\gc, while cyclotron-dominated shocks in
high-field polars may be buried for much lower $\dot m$. To be sure,
the chance for the upward directed radiation to escape will depend on
the details of the geometry.

\subsection{Emerging spectra as a function of $\dot m$}

We have employed a heuristic model to estimate the effect of buried
shocks on the emerging spectra. Specifically, we use a field strength
of $B=14$\,MG (appropriate for AM Herculis) and consider radiation
emerging at an angle against the radial direction of 60$^\circ$. We
consider subcolumns with $\dot m$ values ranging from $10^{-3}$ to
100\,\gcs, disregard the radiative interaction of the subsolumns,
adopt a warm absorber which leaves soft X-rays largely unaffected, and
approximately account for photoelectric absorption in the surrounding
atmosphere using a ray tracing technique. Radiation intercepted by the
white dwarf atmosphere is assumed to be reprocessed into UV radiation
and soft X-rays with a temperature at the base of the individual
subcolumns and their surroundings, which depends on the energy flux
incident on the white dwarf. For simplicity, we assume blackbodies for
the reprocessed flux. This assumption may be acceptable for the hot
sections of the atmosphere, but the cooler UV emitting parts of the
heated pole cap would correctly show the Lyman edge in absorption
(G\"ansicke et al. 1998, Fig. 12). In Fig.\,5, the original emission
and the reprocessed spectra are shown as dashed and solid curves,
respectively. For low $\dot m$, the shock stands above the atmosphere
and the emitted radiation escapes more or less freely, for larger
$\dot m$ the originally emitted radiation is increasingly absorbed and
reprocessed into soft X-rays, until at $\dot m = 100$\,\gcs\ the
emission region is hidden behind an average column density of $\sim
10^{25}$\,\gc\ and is completely reprocessed.

\begin{figure}
\plotfiddle{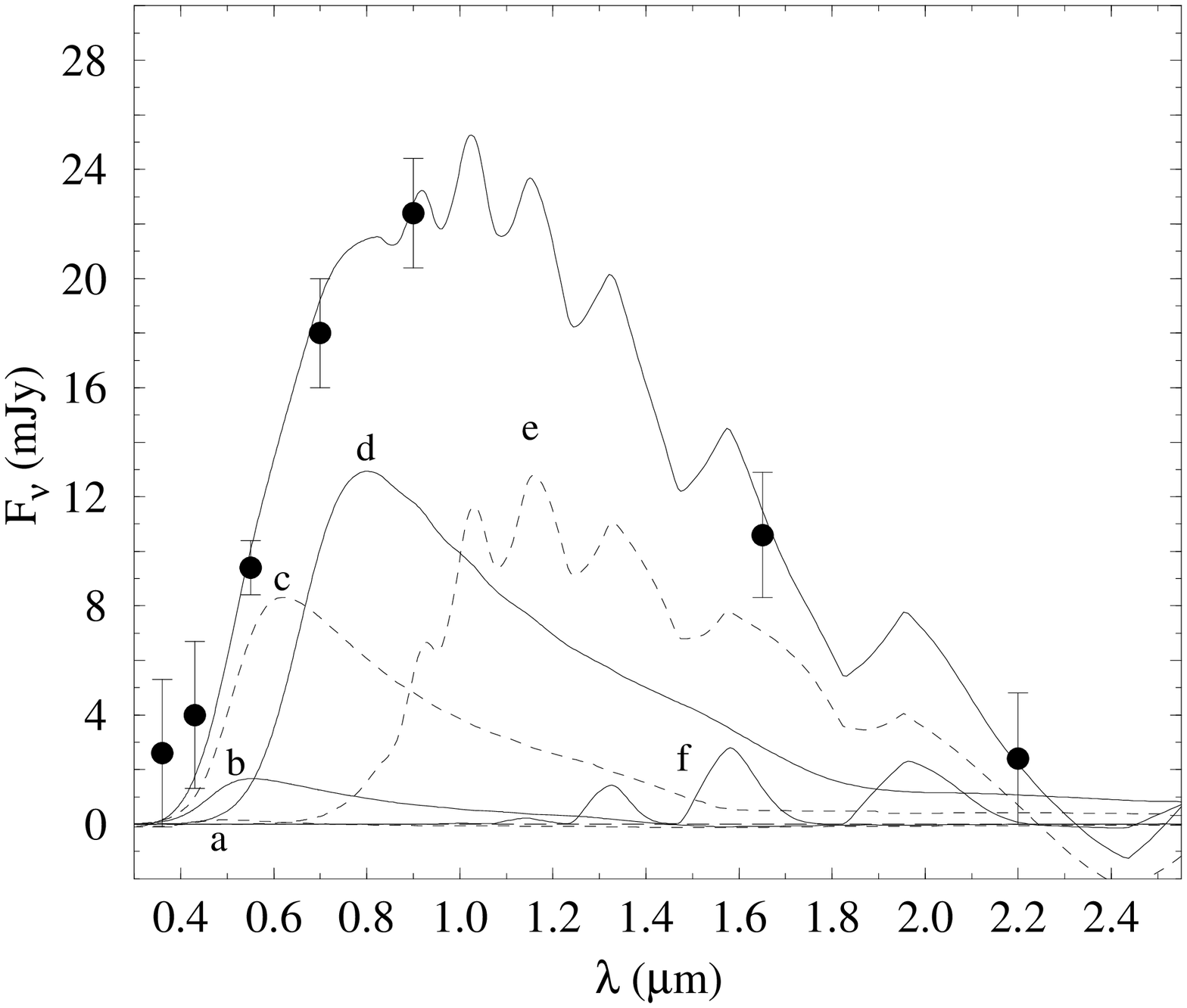}{24mm}{0}{31}{31}{-205pt}{-99pt}
\plotfiddle{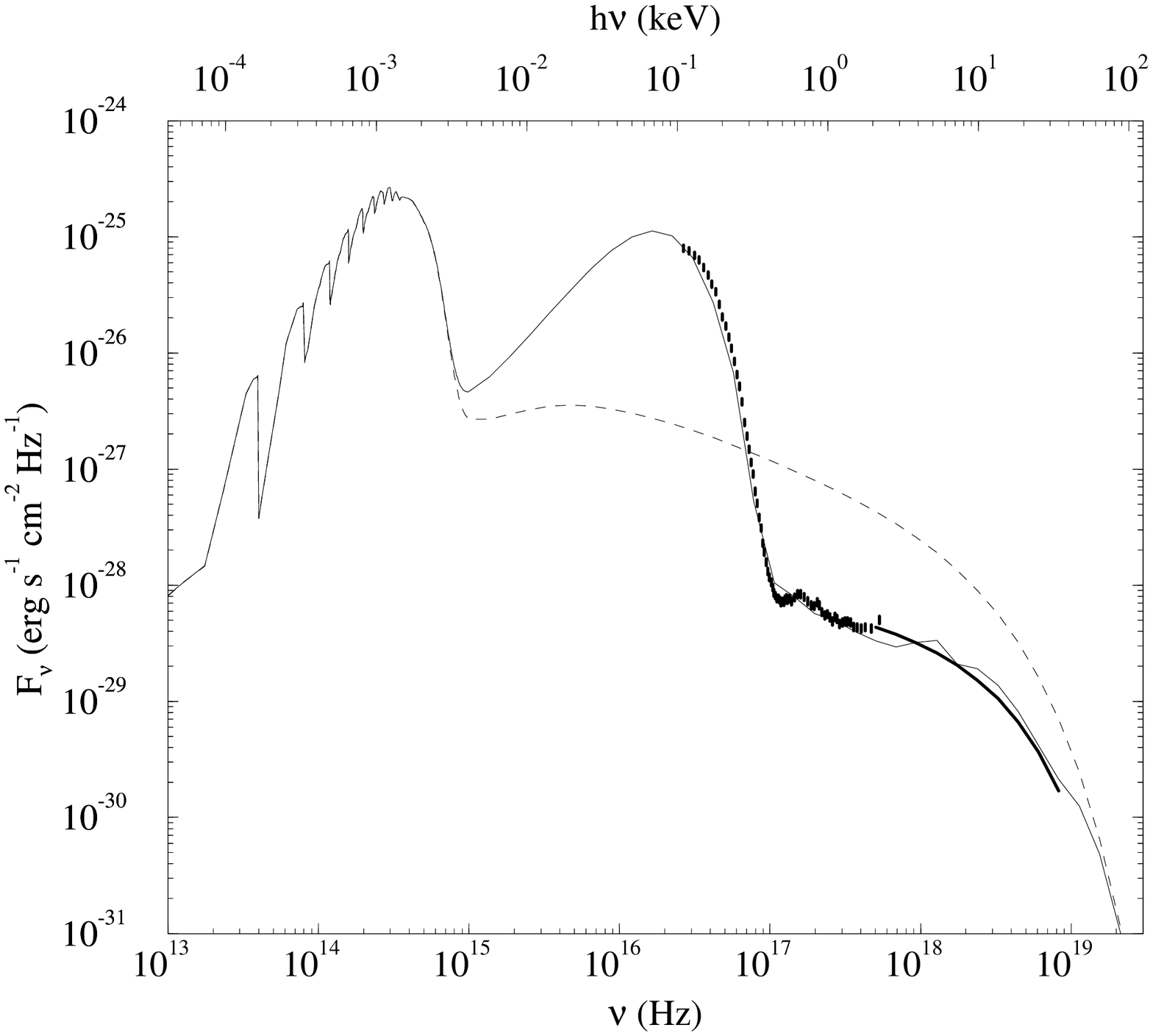}{24mm}{0}{31}{31}{-15pt}{-18pt}
\caption{{\it (a) Left: } Synthesis of optical/IR cyclotron fluxes of
AM Her in its high state (Priedhorsky et al. 1978). {\it (b) Right: }
Synthesis of ROSAT and HEAO-1 X-ray spectrum (G\"ansicke et al. 1995,
Rothschild et al. 1981) (from Beuermann \& Fischer in preparation).}
\end{figure}

\subsection{Spectral synthesis of AM Herculis}

The characteristic shapes of the spectra in Fig.\,5 depend on $\dot m$
and $B$. For a given field strength, such set of spectra can be used
to synthesize the observed spectral energy distribution and thus yield
the generating $\dot m$-distribution. Fig.\,6 shows the result for AM
Herculis, based on the cyclotron and X-ray spectral energy
distributions observed near the respective orbital maxima. The
cyclotron flux (Priedhorsky et al. 1978) is modelled by contributions
with 100\,\gcs\ ( curve a) to $\dot m = 10^{-3}$\,\gcs\ (curve f),
with very small contributions for $\dot m \ga 3$\,\gcs. The hard X-ray
spectrum, on the other hand, originates mostly from mass fluxes around
1~\gcs, and the large soft X-ray flux is produced by $\dot m \ga
10$\,\gcs. Hence, the origin of the individual contributions is to a
sufficient extent mutually exclusive. It is comforting and adds to the
credibility of the result that the contributions for $\dot m$ around
1~\gcs\ derived from the cyclotron and X-ray fluxes are consistent
with each other. The same analysis has been performed for the low
state of AM Herculis, using the infrared cyclotron spectrum of Bailey et
al. (1991) and the low-state ROSAT PSPC spectrum of September 1991.

\subsection{Distribution of Mass Fluxes in the Spot on AM Herculis}

Figure\,7 depicts the contributions $\Delta \dot M$ to the total
time-averaged $\dot M$ in AM Her in its high state as derived from the
spectra in Fig.\,6.  Also shown is the result of a corresponding
analysis of the low state. While the errors in the derived values of
$\Delta \dot M$ are much larger in the low state, a plausible picture
emerges. The high state is characterized by a large contribution from
high $\dot m$ which are largely missing in the low state, while a
drizzle of low $\dot m$ seems to be present at all times (including
most low states observed so far).
Quite possibly, these differences exist already in the flow passing
through the nozzle in L$_1$.
Note that ``100\,\gcs'' is actually a synonym for dense blobs since we
are not able to distinguish between more or less dense blobs. The fate
of blobs of different density is not considered here as long as the
shocks are likely to be buried (compare Frank et al. 1988).

\begin{figure} 
\plotfiddle{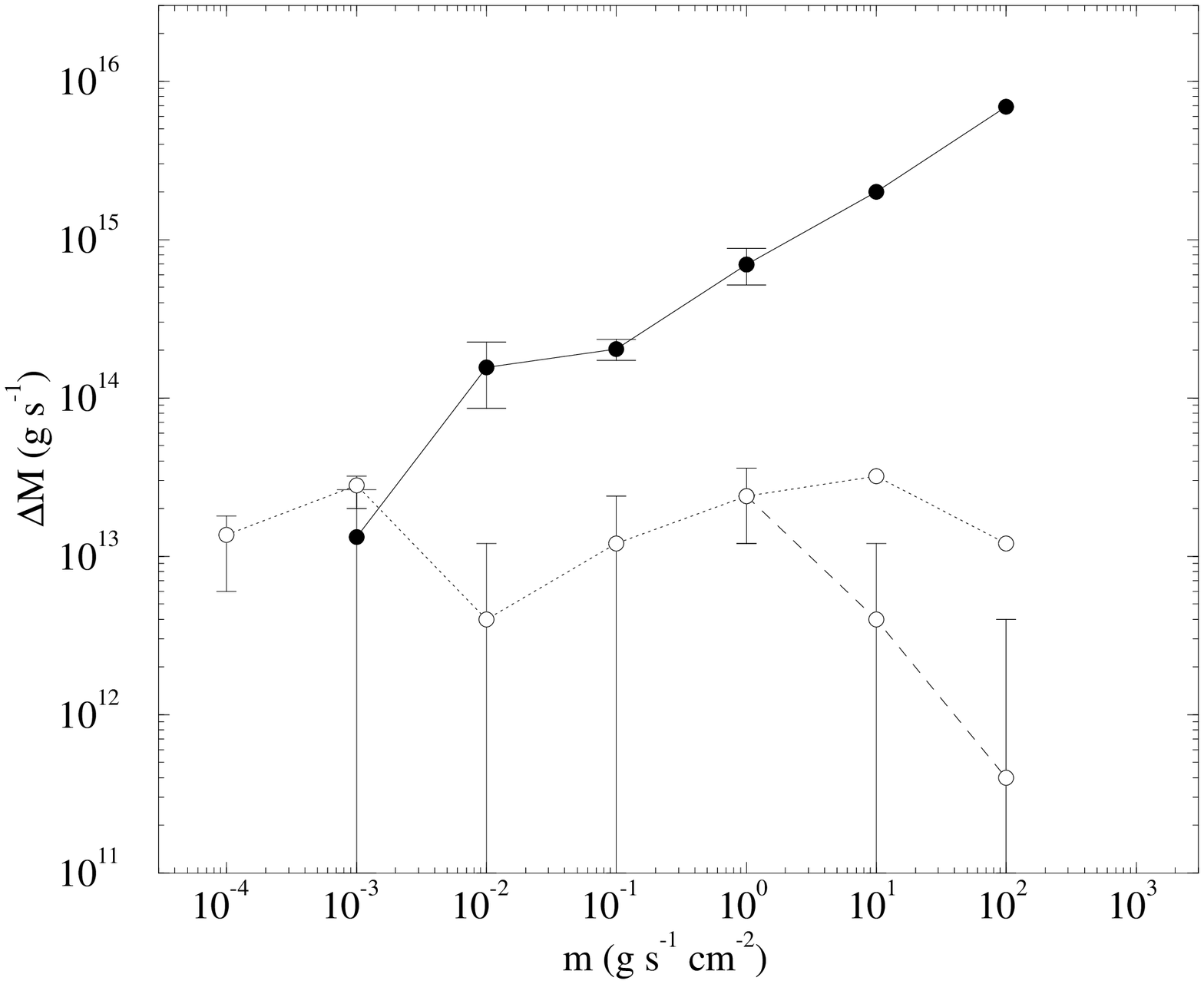}{53mm}{0}{35}{35}{-130pt}{-20pt}
\caption{Contributions $\Delta \dot M$ to the accretion rate $\dot M$
from mass fluxes $\dot m$ between $10^{-4}$ and 100\,\gcs\ for AM
Herculis in its high state (filled circles) and its low state (open
circles) (Beuermann \& Fischer in preparation).}
\end{figure}

\subsection{Variation of Hard vs. Soft X-rax Fluxes}

It is long known that soft X-ray emission does not prevail in all AM
Her stars, a notable exception being some systems with a low polar
field strength (Beuermann \& Schwope 1994, Ramsay et al. 1994, Ramsay
this conference). One possible explanation is the shift of the whole
$\dot m$-distribution towards smaller $\dot m$, which is expected to
occur in systems with lower field strengths \mbox{(Beuermann 1998)}. 

\section{Outlook}

An attempt to model the entire internal structure of the emission
spots in polars is probably futile. Some important aspects of the
relevant physics need additional attention, however. Among them are
the possible heating of electrons by plasma instabilities in mCV
shocks, the effect of the optically thick radiative transfer on
$T_{\rm e}$, and the further development of two-fluid
models. Pioneering studies of two-dimensional radiative transfer may
be enlightening, too. On the observational side, studies high time
resolution with the large ground-based and space-borne telescopes will
undoubtedly provide a rapidly increasing insight into the dynamics and
the spatial structure of the cyclotron as well as the X-ray emission
regions.

\end{document}